\documentclass[10pt, conference, letterpaper]{IEEEtran}
\IEEEoverridecommandlockouts
\usepackage{cite}
\usepackage{amsmath,amssymb,amsfonts}
\usepackage{algorithm2e}
\usepackage{graphicx}
\usepackage[caption=false,font=footnotesize]{subfig}
\usepackage{textcomp}
\usepackage{xcolor}
\def\BibTeX{{\rm B\kern-.05em{\sc i\kern-.025em b}\kern-.08em
    T\kern-.1667em\lower.7ex\hbox{E}\kern-.125emX}}
\usepackage{tabularx}
\usepackage{multirow}
\usepackage[binary-units]{siunitx} 
\DeclareSIUnit{\bps}{bps}
\usepackage{mathrsfs} 

\begin{document}

\title{Beam-Based Multiple Access for IRS-Aided Millimeter-Wave and Terahertz Communications\\
{\footnotesize}
}
\author{\IEEEauthorblockN{Wei Jiang}
\IEEEauthorblockA{Intelligent Networking Research Group\\German Research Center for Artificial Intelligence (DFKI)\\
Kaiserslautern, Germany \\
\and
\IEEEauthorblockN{Hans Dieter Schotten}
\IEEEauthorblockA{Department of Electrical and Computer Engineering\\Rheinland-Pf\"alzische Technische Universit\"at\\ Kaiserslautern-Landau  (RPTU)\\
Kaiserslautern, Germany\\
}
} }
\maketitle

\begin{abstract}
Recently, intelligent reflecting surface (IRS)-aided millimeter-wave (mmWave) and terahertz (THz) communications are considered in the wireless community.  This paper aims to design a beam-based multiple-access strategy for this new paradigm. Its key idea is to make use of multiple sub-arrays over a hybrid digital-analog array to form independent beams, each of which is steered towards the desired direction to mitigate inter-user interference and suppress unwanted signal reflection. The proposed scheme combines the advantages of both orthogonal multiple access (i.e., no inter-user interference) and non-orthogonal multiple access (i.e., full time-frequency resource use). Consequently, it can substantially boost the system capacity, as verified by Monte-Carlo simulations.    
\end{abstract}

\section{Introduction}

Despite the great potential of providing abundant spectral resources for Six-Generation (6G) and beyond systems \cite{Ref_jiang2023full}, millimeter-wave (mmWave) and terahertz (THz) signals suffer from extremely worse propagation conditions raised by high path loss, atmospheric absorption, and weather attenuation, leading to short transmission distances \cite{Ref_jiang2023terahertz}. 
Therefore, a large-scale antenna array with high beamforming gain is generally employed to compensate for this propagation loss. \textit{Digital beamforming} over a large-scale array raises high hardware and energy costs because it requires massive radio frequency (RF) components. \textit{Analog beamforming} provides a low-cost, low-complexity alternative by using a single RF chain, but it suffers from poor performance and hardware constraints. Accordingly, \textit{hybrid beamforming} \cite{Ref_jiang2022initial_ICC} is widely recognized as the most suitable structure for implementing mmWave and THz transceivers. Using a few RF chains and a phase-shifter network, it achieves high performance comparable to digital beamforming with much lower hardware complexity. 

Because of poor diffraction and scattering ability, high-frequency signals heavily rely on line-of-sight (LOS) transmission, which is susceptible to obstacles \cite{Ref_jiang2024terahertz}. The pencil beams formed by a large-scale array further aggravate the signal blockage. This fatal problem severely impacts communications reliability and hinders the practical roll-out of mmWave and THz systems. Recently, a new paradigm of intelligent reflecting surface (IRS)-aided mmWave and THz communications \cite{Ref_ning2021terahertz} is considered to address the above issues by jointly leveraging hybrid beamforming at a base station (BS) and passive reflection at the IRS \cite{Ref_jiang2022dualbeam}. 

In contrast to the conventional system, the use of IRS brings some fundamental particularities in coordinating multi-user signal transmission. Due to the lack of frequency-selective IRS reflection, the widely adopted technique, i.e., orthogonal frequency-division multiple access (OFDMA), is not efficient anymore \cite{Ref_jiang2023userscheduling}. In \cite{Ref_jiang2023capacity, Ref_zheng2020intelligent_COML, Ref_jiang2023orthogonal}, the researchers studied non-orthogonal multiple access (NOMA) for IRS-aided systems. The work in \cite{Ref_jiang2022multiuser} studied the effect of discrete phase shifts on multi-user IRS communications.  The authors of this paper have also proposed and compared several IRS multiple-access schemes in \cite{Ref_jiang2022intelligent, Ref_jiang2023userselection, Ref_jiang2023opportunistic, Ref_jiang2023simple}. 
To the best knowledge of the authors, an appropriate multiple-access method for IRS-aided systems in high-frequency bands is still missing.

Responding to this, we propose a beam-based strategy, coined beam-division multiple access (BDMA), for IRS-aided mmWave and THz communications. This strategy makes full use of multiple sub-arrays over a hybrid digital-analog antenna array to form independent beams, each of which is steered towards the desired direction to mitigate inter-user interference and avoid unwanted signal reflection on the IRS. We design a hierarchical frame structure, where a superframe on the same order of magnitude as large-scale fading deals with angle measurement, while each radio frame conducts scheduling on the granularity of small-scale fading. In addition, user categorizing and grouping are provided, so that the beam-based approach can be combined with time shifting to achieve highly flexible multiple access. The proposed scheme combines the advantages of both orthogonal multiple access (i.e., no inter-user interference) and non-orthogonal multiple access (i.e., full-time-frequency resource use). Consequently, BDMA substantially outperforms FDMA, TDMA, and NOMA in terms of system capacity. The sum rates of different schemes are compared through Monte-Carlo simulations to justify the superiority of the proposed scheme.  



\section{System Model}

\begin{figure}[!t]
    \centering
    \includegraphics[width=0.42\textwidth]{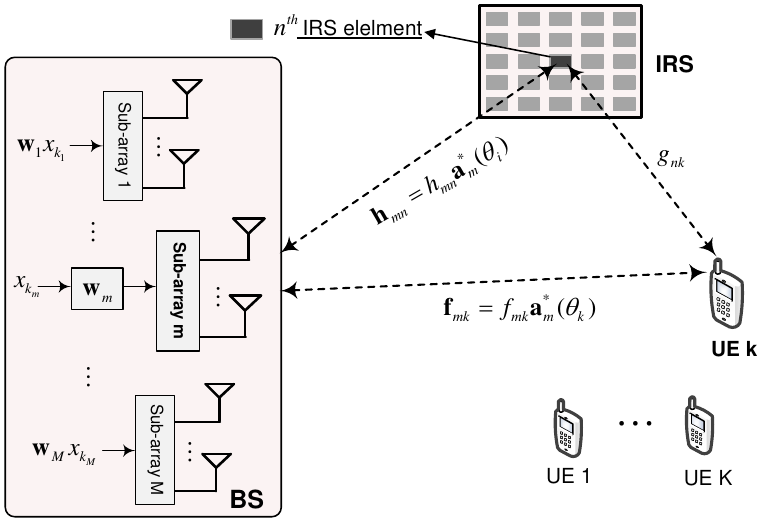}
    \caption{Schematic diagram of an IRS-aided mmWave and THz system.  }
    \label{diagram:system}
\end{figure}
Consider a single-cell massive MIMO communications system operating in mmWave or THz band, which is comprised of a base station, $K$ users, and an IRS, as shown in \figurename \ref{diagram:system}. The BS is equipped with a hybrid digital-analog antenna array, where a large number of $N_b$ antennas are driven by $M$ RF chains with $M\ll N_b$. There are two basic forms of hardware implementation: fully-connected and partially-connected hybrid beamforming. Without losing any generality, this paper focuses on the latter since it lowers complexity and power consumption by substantially reducing the number of analog phase shifters. Correspondingly, the array is divided into $M$ sub-arrays, each of which consists of $N_s=N_b/M$ non-overlapping elements (suppose $N_s$ is an integer) and is driven by one of the RF chains \cite{Ref_jiang2022initial}.

\begin{figure}[!t]
    \centering
    \includegraphics[width=0.4\textwidth]{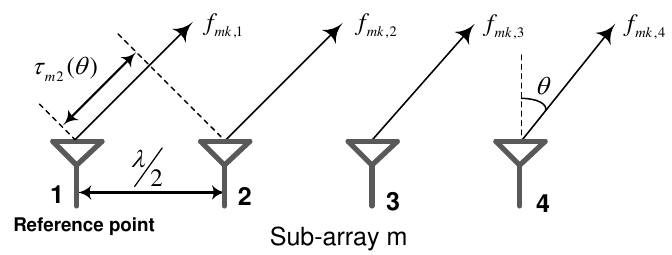}
    \caption{Far-field geometry of sub-array $m$. }
    \label{diagram:ULA}
\end{figure}

The inter-element spacing of a hybrid digital-analog array is usually small, e.g., half wavelength, to facilitate beam steering over  correlated elements. 
In a typical macro-cell scenario, the scatters locate in the surrounding of user equipment (UE). The diameter of the scattering area is sufficient small in comparison with the UE-BS distance. As a result, the angle of departure (AOD) for various signal paths in the far field is approximately identical, denoted by $\theta$, as illustrated in \figurename \ref{diagram:ULA}. Generally, the first antenna of each sub-array is set as the reference point. Denoting the channel coefficient between the reference point of sub-array $m$, $m=1,2,\ldots,M$ and UE $k$, $k=1,2,\ldots,K$ by $f_{mk,1}$, the channel coefficient between a typical element $n_s$ of this sub-array and UE $k$ is expressed as \cite{Ref_yang2013random}
\begin{equation} 
    f_{mk,n_s} = f_{mk,1}e^{-j2\pi f_c\tau_{mn_s}(\theta)},
\end{equation} where $\tau_{mn_s}(\theta)$ denotes the difference of propagation time between the reference point and the $n_s^{th}$ element, which is a function of $\theta$,  and $f_c$ represents the carrier frequency.
Subsequently, the channel vector between sub-array $m$ and UE $k$ is given by
\begin{align} \nonumber \label{EQN_CHVector}
    \mathbf{f}_{mk}&=\Bigl[f_{mk,1},f_{mk,2},\ldots,f_{mk,N_s}\Bigr]^T\\
    &=f_{mk,1}\Bigl[1,e^{-j2\pi f_c\tau_{m2}(\theta)},\ldots,e^{-j2\pi f_c\tau_{mN_s}(\theta)}\Bigr]^T.
\end{align}
We write 
\begin{equation}
    \textbf{a}_m(\theta) = \left[1,e^{j2\pi f_c\tau_{m2}(\theta)},\ldots,e^{j2\pi f_c\tau_{mN_s}(\theta)}   \right]^T
\end{equation} to denote the steering vector of sub-array $m$ and let $f_{mk}=f_{mk,1}$ for simplicity. Then, \eqref{EQN_CHVector} is rewritten as  $\mathbf{f}_{mk}=f_{mk}\mathbf{a}_m^*(\theta)$.
Thus, we can obtain the overall channel vector between the BS and UE $k$ as
\begin{align} \nonumber 
    \mathbf{f}_k&=\Bigl[\mathbf{f}_{1k}^T,\mathbf{f}_{2k}^T,\ldots,\mathbf{f}_{Mk}^T\Bigr]^T\\
    &=\Bigl[f_{1k}\mathbf{a}_1^H(\theta),f_{2k}\mathbf{a}_2^H(\theta),\ldots,f_{Mk}\mathbf{a}_M^H(\theta)  \Bigr]^T.
\end{align}

The users are categorized into two types: those locating in the cell center are called \textit{the near users (NUs)}, and those who are far away from the BS or blocked by obstacles are referred to as \textit{the far users (FUs)}. To aid the far users that suffer from poor signal quality, a surface is deployed at the cell edge or a particular location that can bypass the blockage. An IRS contains $N$ reflecting elements, each of which is mathematically modeled by a reflection coefficient $c_n=\alpha_n e^{j\phi_n}$, with a phase shift $\phi_n\in [0,2\pi)$ and attenuation amplitude $\alpha_n\in [0,1]$. We generally set $\alpha_n=1$, $\forall n=1,2,\ldots,N$ to enlarge the received signal strength and lower the complexity of hardware implementation. Thus, only phase shifts needed to be dynamically adjusted by a smart controller. 

Similarly, the channel vector between the $m^{th}$ sub-array and the $n^{th}$ reflecting element can be expressed as
\begin{equation} \label{EQN_GC_hmn}
\mathbf{h}_{mn}=h_{mn}\mathbf{a}_m^*(\theta),
\end{equation} 
where $h_{mn}$ denotes the channel coefficient between the reference point of sub-array $m$ and the $n^{th}$ IRS element.
The channel from the BS to the $n^{th}$ reflecting element equals 
\begin{align} \nonumber 
    \mathbf{h}_n&=\Bigl[\mathbf{h}_{1n}^T,\mathbf{h}_{2n}^T,\ldots,\mathbf{h}_{Mn}^T\Bigr]^T\in \mathbb{C}^{N_b\times 1}\\
    &=\Bigl[h_{1n}\mathbf{a}_1^H(\theta),h_{2n}\mathbf{a}_2^H(\theta),\ldots,h_{Mn}\mathbf{a}_M^H(\theta)  \Bigr]^T,
\end{align} and the BS-IRS channel matrix is $\mathbf{H}\in \mathbb{C}^{N\times N_b}$, where the $n^{th}$ row of $\mathbf{H}$ equals to $\mathbf{h}_n^T$. 
In addition, the link from the $n^{th}$ reflecting element to the $k^{th}$ UE is a single-input single-output channel, which is denoted by $g_{nk}$, $\forall n=1,\ldots,N$ and $\forall k=1,\ldots,K$. The channel between the IRS and UE $k$ can be expressed as an $N\times 1$ vector $\mathbf{g}_{k}=\Bigl[g_{1k},g_{2k},\ldots,g_{Nk}\Bigr]^T$.


\section{Beam-Division Multiple Access}
\begin{figure}[!t]
    \centering
    \includegraphics[width=0.43\textwidth]{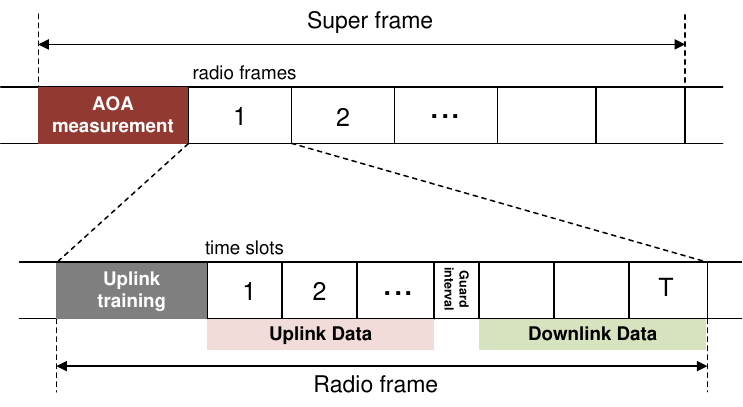}
    \caption{The proposed hierarchical frame structure.}
    \label{diagram:Frame}
\end{figure}

\begin{figure}[!t]
    \centering
    \includegraphics[width=0.36\textwidth]{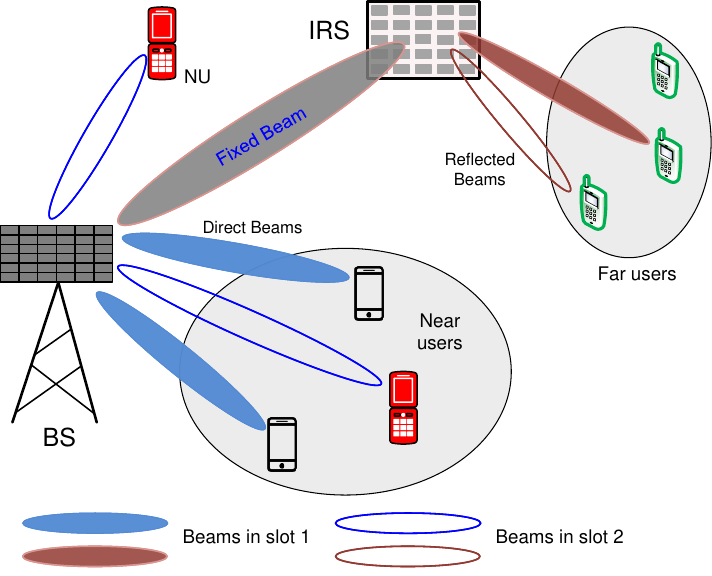}
    \caption{Illustration of BDMA for an IRS-aided mmWave and THz system. The signal transmission of the far users are aided by the IRS, while the near users directly access the BS. Without losing generality, this figure only shows the time shifting for two slots and the user grouping with three active users (one FU and two NUs) per slot.}
    \label{diagram:SDMA}
\end{figure}

The signal transmission of each TDMA or FDMA user only exploits $1/K$ of the total time-frequency resource. This is inefficient according to the Shannon theorem. Although NOMA users can fully use the resource, strong inter-user interference constraints the performance. To enable an efficient multi-user IRS-aided mmWave and THz communications, we propose a beam-based multiple access scheme in this section. The key idea is to make full use of multiple sub-arrays over a hybrid digital-analog array to form independent beams, each of which is steered towards the desired direction to mitigate inter-user interference and avoid unwanted signal reflection. BDMA combines the advantages of both orthogonal multiple access (i.e., no inter-user interference) and non-orthogonal multiple access (i.e., full time-frequency resource utilization).
The proposed scheme is depicted as follows:
\begin{itemize}
    \item \textit{User categorizing:} Classify the set of all users $\mathscr{U}=\{1,2,\cdots,K\}$ into two subsets: the far users $\mathscr{U}_f$ and the near users $\mathscr{U}_n$, subjecting to $\mathscr{U}_f \bigcup \mathscr{U}_n=\mathscr{U}$ and $\mathscr{U}_f \bigcap \mathscr{U}_{n}=\varnothing$. The philosophy is as follows: any NU can directly access to the BS because of its good signal quality, whereas an FU with weak signal strength needs the aid of the IRS. The categorization can be conducted based on averaged power strength, instantaneous channel fading, or their location information. 
    \item \textit{Angle-of-Arrival (AOA) measurement per super frame:} Since the IRS is fixed, its angle $\theta_i$ relative to the BS is deterministic. The BS only needs to periodically measure the AOA of users $\theta_k$, $\forall k\in \mathscr{U}$, which can be efficiently estimated through classical algorithms, such as MUSIC and ESPRIT \cite{Ref_MUSIC}. Compared with small-scale fading, the angle of a user vary slowly on the same order of magnitude as large-scaling fading. As illustrated in \figurename \ref{diagram:Frame}, we design the structure of a super frame, within which the angle change of each user is negligible. AOA measurement is conducted at the header and the measures are used for the whole super frame. In a massive MIMO system, time-division duplexing (TDD) mode is usually adopted to simplify channel estimation. Due to channel reciprocity, the AOD in the downlink is assumed to be the same as the measured AOA in the uplink.
    \item \textit{Channel state information (CSI) acquisition per radio frame:} A super frame contains a number of radio frames, each of which is comprised of three main parts: the uplink training, uplink data transmission, and downlink data transmission.  At each radio frame, the BS first gets CSI estimates $\mathbf{H}$, $\mathbf{f}_k$, and $\mathbf{g}_k$, $\forall k$,  through effective channel estimation methods such as \cite{Ref_you2020channel}. Small-scale fading is assumed to keep constant for the whole radio frame. 
    \item \textit{User grouping:} A radio frame is further divided into $T$ time slots. Each slot can maximally serve $M$ users in a spatial-division manner since $M$ sub-arrays form $M$ beams simultaneously, as shown in \figurename \ref{diagram:SDMA}. The IRS can be fully optimized for a particular user because each IRS element induces only one phase shift at a particular instant. Hence, we select one FU\footnote{If the system is equipped with multiple IRS surfaces, the BS can serve multiple far users simultaneously, one FU per IRS.}, and $M-1$ NUs for each slot.
    The implementation of selection strategies is flexible, e.g., round-robin, proportional fairness, or priority-based for ultra-reliability, low-latency users. 
    \item \textit{IRS-aided transmission for the far user:} Without losing generality, we assign the first sub-array for the FU.  This sub-array maximizes its beam gain in the direction of $\theta_i$, regardless of different slots or radio frames. At slot $t$, the IRS  is tuned to the optimal phase shifts $\phi_n^\star[t]$ accordingly to achieve coherent combining at the FU.  
    \item \textit{Direct access for the near users:} Meanwhile, other $M-1$ sub-arrays serve $M-1$ near users simultaneously, where sub-array $m$, $m=2,\ldots,M$ forms a direct beam towards its assigned user.
    \item \textit{Time shifting:} Once the completion of a time slot, another user group, consisting of the same or  different users, gets service in the next slot. 
\end{itemize}

\section{Optimization Design of BDMA} 
This section further elaborates the BDMA scheme by discussing the design of optimization parameters. 
Since hybrid beamforming needs only a few RF chains, the number of users $K$ is usually greater than $M$, i.e., $K>M$. At each slot, the  user group has $M$ active users selected from a total of $K$ users. For ease of illustration, we write $k_m$ to denote the user served by sub-array $m$, and the user group is denoted by $\mathscr{U}_{ts}=\{k_1,k_2,\ldots,k_M\}$. As illustrated in \figurename \ref{diagram:system}, the $m^{th}$ sub-array independently forms a beam to transmit $x_{k_m}$ intended for UE $k_m$, where $\mathbb{E}\left[|x_{k_m}|^2\right]=1$.

Each antenna imposes a weighting coefficient on its carried signal by using a analog shifter to induce a phase shift, e.g., $\psi_{mn_s}$ on the $n_s^{th}$ element of the $m^{th}$ sub-array. Thus, the weighting vector of sub-array $m$ can be expressed by
\begin{equation}
    \textbf{w}_{m}= \left[e^{j\psi_{m1}}, e^{j\psi_{m2}},\ldots, e^{j\psi_{mN_s}}   \right]^T,
\end{equation} corresponding to a beam pattern 
\begin{equation} \label{eqn:ICC:beampattern}
   B_m(\theta)= \textbf{a}_m^H(\theta)\textbf{w}_m=\sum_{n=1}^{N_s}  e^{-j2\pi f_c\tau_{mn_s}(\theta)}e^{j\psi_{mn_s}}. 
\end{equation} 
Note that $B_m(\theta)$ is a function of $\theta\in [0,2\pi)$, whereas $B_m(\theta_i)$ is a scalar denoting the pattern gain in the direction of the IRS, and $B_m(\theta_k)$ is also a scalar in a particular angle of $\theta_k$. For ease of notation, we hereinafter denote 
\begin{equation}
    b_{mi} = B_m(\theta_i), \:\:\:\:\: b_{mk} = B_m(\theta_k).
\end{equation}
A typical user $k$ observes the received signal:
\begin{align} \nonumber \label{IRSEQN:downlinkdiscreteModel}
    y_k&=\sqrt{\frac{P_d}{M}}  \sum_{n=1}^N g_{nk} e^{j\phi_n} \sum_{m=1}^M h_{mn}b_{mi}  x_{k_m} \\
    &+ \sqrt{\frac{P_d}{M}} \sum_{m=1}^M f_{mk}b_{mk}   x_{k_m} + z_k\\ \nonumber
    &=\sqrt{\frac{P_d}{M}} \sum_{m=1}^M\Biggl( \sum_{n=1}^N g_{nk} e^{j\phi_n}  h_{mn}b_{mi} + f_{mk}b_{mk} \Biggr)  x_{k_m} + z_k.
\end{align}

Substituting $k=k_m$ into \eqref{IRSEQN:downlinkdiscreteModel}, we can analyze the received signal for a particular user $k_m$, as given by \eqref{IRSEQN:downlink_MUI} on the top of the next page,
\begin{figure*}[!t]
\setcounter{equation}{27}
\begin{equation}  \label{IRSEQN:downlink_MUI}
    y_{k_m}=\underbrace{\sqrt{\frac{P_d}{M}}\Biggl( \sum_{n=1}^N g_{n{k_m}} e^{j\phi_n}  h_{mn}b_{mi} + f_{m{k_m}}b_{m{k_m}} \Biggr)  x_{k_m}}_{\text{Desired\:Signal:}\:\mathcal{D}_{k_m}}+\underbrace{ \sqrt{\frac{P_d}{M}}\sum_{m'\neq m}^M\Biggl( \sum_{n=1}^N g_{nk_m} e^{j\phi_n}  h_{m'n}b_{m'i} + f_{m'k_m}b_{m'k_m} \Biggr)  x_{k_{m'}}}_{\text{Multi-User\:Interference:}\:\mathcal{I}} + z_{k_m},
\end{equation}
\end{figure*} which is comprised of three terms: the desired signal, multi-user interference, and white noise.
The aim of the multi-access algorithm design is to maximize the spectral efficiency of a typical user through optimally selecting weighting vectors and reflecting coefficients.
Generally, we can either maximize the strength of the desired signal (as maximal-ratio combining and matched filtering) 
\begin{equation}  \setcounter{equation}{27}
    |\mathcal{D}_{k_m}|^2 = \frac{P_d \Bigl|\sum_{n=1}^N g_{nk_m} e^{j\phi_n} h_{mn}  b_{mi}  + f_{mk_m} b_{mk}\Bigr|^2 }{M \sigma_z^2}.
\end{equation}
or minimizing the multi-user interference $ |\mathcal{I}|^2 $ (like zero forcing). In this paper, the strategy is to maximize the desired signal power by steering the beam towards the desired direction, while trying to avoid the main beams of other users. 
 
\subsection{Reflected Beam Optimization}
First, let's look at the optimization of the reflected beam towards the IRS. Without losing generality, the first sub-array is dedicated to the selected far user. The maximization of $ |\mathcal{D}_{k_1}|^2 $ results in the following optimization problem
\setcounter{equation}{28}
\begin{equation}  \label{IRSoptimizationPrlbme}  
\begin{aligned} 
\max_{\boldsymbol{\Phi},\mathbf{w}_1}\quad &  \biggl|\sum_{n=1}^N g_{nk_1} e^{j\phi_n} h_{1n}  b_{1i}  + f_{1k_1} b_{1k_1}\biggr|^2\\
\textrm{s.t.} \quad & \|\mathbf{w}_{1}\|^2\leqslant 1\\
  \quad & \phi_n\in [0,2\pi), \: \forall n=1,2,\ldots,N. 
\end{aligned}
\end{equation} To maximize the beam gain towards the IRS, we need to solve
\begin{equation}  
\begin{aligned} 
\max_{\mathbf{w}_1} \quad & \Bigl|\textbf{a}_1^H(\theta_i)  \mathbf{w}_1\Bigl|^2   \\
\textrm{s.t.} \quad & \|\mathbf{w}_1\|^2\leqslant 1. 
\end{aligned}
\end{equation}
Since the locations of the BS and IRS are fixed, $\theta_i$ is easy to know and keeps constant in a long-term basis.   The optimal weighting vector is computed by $\mathbf{w}_1^\star=\sqrt{1/N_s}\textbf{a}_1(\theta_i)$, resulting in the maximal gain
\begin{equation} \label{optimalWeight1}
    b_{1i}=B_1(\theta_i)=\sqrt{\frac{1}{N_s}} \textbf{a}_1^H(\theta_i)  \textbf{a}_1(\theta_i)=\sqrt{N_s}.
\end{equation} 
In addition to the main beam targeting the IRS, the sidelobe gain in the direction of the far user $\theta_{k_1}$ can be computed by
\begin{equation}
    b_{1k_1} =B_1(\theta_{k_1})=\sqrt{\frac{1}{N_s}} \textbf{a}_1^H(\theta_{k_1})  \textbf{a}_1(\theta_i).
\end{equation}
If the number of elements per sub-array is sufficiently large, the energy can be concentrated in a very small beamwidth, where the sidelobe is remarkably weak compared with the main beam. For simplicity, we assume that the sidelobe of any beam pattern is negligible, e.g., $|b_{1k_1}|=0$.
Thus, the optimization problem in \eqref{IRSoptimizationPrlbme} is reduced to
\begin{equation}  
\begin{aligned} 
\max_{\boldsymbol{\Phi}}\quad &  \biggl|\sqrt{N_s} \sum_{n=1}^N g_{nk_1} e^{j\phi_n} h_{1n} \biggr|^2\\
\textrm{s.t.} 
  \quad & \phi_n\in [0,2\pi), \: \forall n=1,2,\ldots,N, 
\end{aligned}
\end{equation}
which is equivalent to an IRS system with a single-antenna BS. 
Therefore, the optimal phase shift for reflecting element $n$ equals to 
\begin{equation} \label{IRSeqn:optimalphase}    
\phi_n^\star = \mod \Bigl[\psi_a-\arg(h_{1n})-\arg(g_{nk_1}), 2\pi   \Bigr],
\end{equation}
where $\psi_a$ stands for an arbitrary phase value. It implies that the phase shift of each reflected signal is compensated, such that the residual phase of each branch is equal to $\psi_a$, for the purpose of coherent combining at the receiver.

\subsection{Direct Beam Optimization}
In addition to the FU $k_1$, the remaining $M-1$ users in the user group are NUs, which directly access to the BS. Sub-arrays $m=2,3,\ldots,M$ generates $M-1$ independent beams for these near users. For a typical NU $k_m\in\{k_2,\ldots,k_M\}$, the maximization of $|\mathcal{D}_{k_m}|^2$ is achieved by,  as \eqref{optimalWeight1}, using the optimal weighting vector 
\begin{equation} \label{optimalWeight2}
    \mathbf{w}_{m}^\star=\sqrt{\frac{1}{N_s}}\textbf{a}_{m}(\theta_{k_m}). 
\end{equation} Thus, the direct beam gain is maximized to 
\begin{equation}
    b_{mk_m}=B_{m}(\theta_{k_m})=\sqrt{\frac{1}{N_s}} \textbf{a}_m^H(\theta_{k_m})  \textbf{a}_m(\theta_{k_m})=\sqrt{N_s}.
\end{equation} 

With the outcome of the optimization parameters, the proposed BDMA scheme is summarized as \algorithmcfname \ref{alg:IRS001}. 
\SetKwComment{Comment}{/* }{ */}
\RestyleAlgo{ruled}
\begin{algorithm} 
\caption{Beam-Division Multiple Access} \label{alg:IRS001}
\SetKwInOut{Input}{input}\SetKwInOut{Output}{output} \SetKwInput{kwInit}{Initialization}
\Input{steering vectors $\mathbf{a}_m(\theta)$, $\forall m=1,2,\ldots,M$}
\Input{AOA of the IRS $\theta_i$}
\kwInit {$\mathbf{w}_1^\star\gets\frac{1}{\sqrt{N_s}}\mathbf{a}_1(\theta_i)$
}
\ForEach{super frame}
{estimate $\theta_k$, $\forall k=1,2,\ldots,K$\;
classify $\mathscr{U}=\{1,2,\cdots,K\}$ to $\mathscr{U}_f$ and $\mathscr{U}_n$\;}
\ForEach{radio frame}{
  estimate $\mathbf{H}$, $\mathbf{g}_k$, and $\mathbf{f}_k$, $\forall k$\;
  \ForEach{time slot}
 {build a user group $\mathscr{U}_{ts}=\{k_1,k_2,\ldots,k_M\}$\;
 get $\phi_n^\star$ as \eqref{IRSeqn:optimalphase}\; 
  adjust phase shifts using $\phi_n^\star$\;
  $\mathbf{w}_m^\star\gets\sqrt{1/N_s}\textbf{a}_m(\theta_{k_m})$\;
  transmit $\mathbf{w}_1^\star x_{k_1}$ over sub-array 1\;
  transmit $\mathbf{w}_m^\star x_{k_m}$ over sub-array $m$\;
  }
  }
\end{algorithm}

\section{Numerical Results}

In this section, we first explain the simulation setup with detailed parameters and then use some representative numerical results to verify the superiority of the proposed scheme. Cumulative distribution function (CDF) of the sum spectral efficiency of different multiple-access techniques is compared. In addition, we also observe the $95\%$-likely spectral efficiency, which is usually used as the measure of cell-edge user performance and the $50\%$-likely or median spectral efficiency to get an insightful view. We design a particular simulation scenario consisting of a cell-center area and a cell-edge area, referring to \cite{Ref_jiang2022multiuser}.  Without losing generality, the cell center is a square with the side length of \SI{125}{\meter} while the cell edge is another square from $(100\si{\meter},100\si{\meter})$ to $(150\si{\meter},150\si{\meter})$. The FUs at the cell edge suffer from weak signal power, which can also emulate the users under blockage. A hybrid array is installed at the central point $(0,0)$, while an IRS is placed at $(125\si{\meter},125\si{\meter})$ to aid the FUs. The number of users $K$ is intentionally selected to be the same as the number of sub-arrays $M$ to simplify user grouping and time shifting. It is observed that this setting does not affect the performance comparison. Last but not least, the number of reflecting elements is set to $N=200$.

\begin{figure*}[!t]
\centerline{
\subfloat[]{
\includegraphics[width=0.36\textwidth]{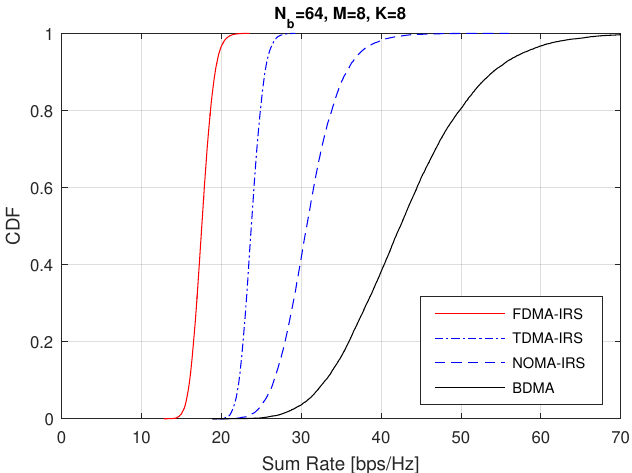}
\label{Fig_results1}
}
\hspace{30mm}
\subfloat[]{
\includegraphics[width=0.36\textwidth]{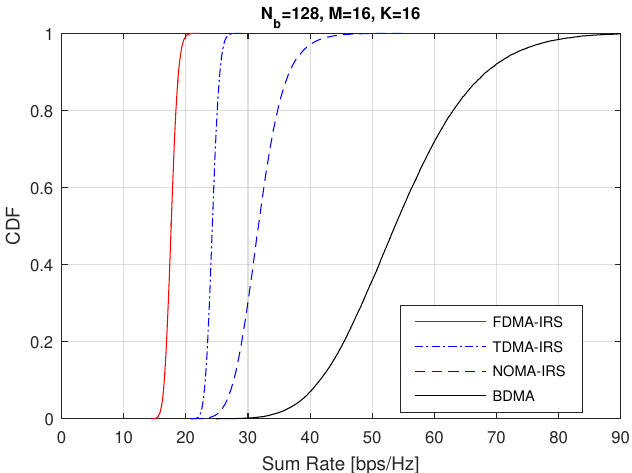}
\label{Fig_results2}
}}
\caption{Performance comparison of different multiple-access schemes in terms of CDFs of sum spectral efficiency: (a) a $64$-antenna base station with $8$ sub-arrays serving $8$ users, and (b) a $128$-antenna base station with $16$ sub-arrays serving $16$ users. }
\label{Fig_Result}
\end{figure*}
Due to the rich scattering around the users, the BS-UE and IRS-UE channels generally follow Rayleigh distribution. Hence, their channel coefficients $f$ and $g$ are circularly symmetric complex Gaussian random variables denoted by $f\sim \mathcal{CN}(0,\sigma_f^2)$ and $g\sim \mathcal{CN}(0,\sigma_g^2)$, respectively. For calculating large-scale fading, we modify the typical COST-Hata model \cite{Ref_jiang2021impactcellfree} to $\sigma^2 [\mathrm{dB}]=-166-35\lg (d)+ \mathcal{X}$,
where the path loss at the reference distance of \SI{1}{\meter} is increased from $-136\mathrm{dB}$ to $-166\mathrm{dB}$, namely an extra loss of $30\mathrm{dB}$ is applied to reflect high path loss, atmospheric absorption, and weather attenuation at mmWave and THz frequencies, $d$ is the propagation distance, and $\mathcal{X}$ stands for shadowing fading with the standard derivation of $8\mathrm{dB}$. In contrast, the BS-IRS channel has a strong LOS path, especially in outdoor environments, because the positions of the IRS is deliberately selected. Therefore, $h$ usually follows Rician distribution with mean $\mu$ and variance $\sigma_h^2$, i.e., $h\sim \mathcal{CN}(\mu,\sigma_h^2)$. Its path loss can be computed through $L(d)=\frac{L_0}{d^{-\alpha}}$,
where $\alpha=2$ means the free-space path loss exponent, and the Rician factor is set to $K=5$. We use a large path loss $L_0=\SI{60}{\decibel}$ at the reference distance of \SI{1}{\meter} to severe propagation conditions at mmWave and THz frequencies.
The BS power constraint is $P_d=20\mathrm{W}$ over a signal bandwidth of $B_w=20\mathrm{MHz}$, conforming with the realistic parameters of 3GPP LTE standard.  The variance of white noise is figured out by $\sigma_z^2= b\cdot B_w\cdot T_0\cdot F_n$ with the Boltzmann constant $b$, temperature $T_0=290 \mathrm{Kelvin}$, and the noise figure  $F_n=9\mathrm{dB}$.

In \figurename \ref{Fig_results1}, we illustrate the performance of an IRS-aided mmWave and THz system with a total of $N_b=64$ BS antennas, which are divided into $M=8$ sub-arrays to serve $K=8$ users. The FDMA scheme achieves the $95\%$-likely spectral efficiency of \SI{15.64}{\bps\per\hertz^{}}, and the $50\%$-likely spectral efficiency of  \SI{17.55}{\bps\per\hertz^{}}. TDMA outperforms FDMA with a sum-rate increase of approximately \SI{6}{\bps\per\hertz^{}}. That is because the IRS is able to provide time-selective reflection particularly optimized for each TDMA user, whereas only the signal transmission of one FDMA user can be aided by the IRS owing to the lack of frequency-selective IRS elements. As expected, NOMA is superior to TDMA, where the $95\%$-likely and $50\%$-likely rates are boosted to approximately \SI{25.5}{\bps\per\hertz^{}} and \SI{30.7}{\bps\per\hertz^{}}, respectively. That is because the signal transmission of each NOMA user fully exploits the whole time-frequency resource. In contrast, TDMA and FDMA merely use $1/K$ of the total resource, leading to a remarkable rate loss according to the Shannon theorem. But the full resource reuse in NOMA causes strong inter-user interference. Although SIC is employed, some interference is still remaining and treated as noise at the signal detection, which negatively affects the performance. The proposed scheme can not only fully exploit the time-frequency resource at signal transmission but also avoid the inter-user interference completely by means of spatial reuse. Hence, BDMA reaps substantial performance improvement with the $95\%$-likely spectral efficiency of around \SI{30.9}{\bps\per\hertz^{}} and the $50\%$-likely spectral efficiency of \SI{42.4}{\bps\per\hertz^{}}.    
Furthermore, we compare these schemes in the simulation setup of $N_b=128$, $M=16$, and $K=16$. As we can see in \figurename \ref{Fig_results2},  similar conclusions can be drawn from the numerical results. 

\section{Conclusions}
This paper proposed a beam-division multiple-access scheme for IRS-aided mmWave and THz communications. Making use of sub-arrays over a hybrid digital-analog array, independent beams are steered towards the desired direction to mitigate inter-user interference and avoid unwanted signal reflection. The signal transmission of each BDMA user fully exploits the time-frequency resource, avoiding the inefficiency of orthogonal multiple access. Compared to non-orthogonal multiple access, BDMA completely suppresses inter-user interference through spatial reuse. Therefore, it can substantially improve the capacity of an IRS-aided mmWave and THz system, as verified by Monte Carlo simulations.





%

\bibliographystyle{IEEEtran}
\bibliography{IEEEabrv,Ref_INFOCOM2022}

\end{document}